\documentclass[10pt,preprint,superscriptaddress,amssymb,nobibnotes,aps,prl,showpacs]{revtex4-1}
\usepackage{graphicx, amsmath, gensymb, textcomp}

\begin{document}
Copyright (2015) American Physical Society. This article may be downloaded for personal use only. Any other use requires prior permission of the author and the American Physical Society.

The following article appeared in Phys. Rev. B 91, 184415 (2015) and may be found at http://link.aps.org/doi/10.1103/PhysRevB.91.184415
\newpage

\title{Excitation of magnetic precession in bismuth iron garnet via a polarization-independent impulsive photo-magnetic effect}

\author{Benny Koene}
\email{b.koene@science.ru.nl}
\affiliation{Radboud University Nijmegen, Institute for Molecules and Materials, Heyendaalseweg 135, 6525~AJ Nijmegen, The Netherlands}
\author{Marwan Deb}
\author{Elena Popova}
\author{Niels Keller}
\affiliation{GEMaC, CNRS-Universit{\'e} de Versailles St.\ Quentin en Yvelines, 45 avenue des Etats-Unis, 78035 Versailles Cedex, France}
\author{Theo Rasing}
\author{Andrei Kirilyuk}
\affiliation{Radboud University Nijmegen, Institute for Molecules and Materials, Heyendaalseweg 135, 6525~AJ Nijmegen, The Netherlands}

\date{\today}

\begin{abstract}
A polarization independent, non-thermal optical effect on the magnetization in bismuth iron garnet is found, in addition to the circular polarization dependent inverse Faraday effect and the linear polarization dependent photo-induced magnetic anisotropy. Its impulsive character is demonstrated by the field dependence of the amplitude of the resulting precession, which cannot be explained by a long living photo or heat induced anisotropy.
\end{abstract}

\pacs{78.47.-p, 75.30.Gw, 75.78.Jp, 75.50.Gg}

\maketitle

Controlling the magnetization dynamics with femtosecond laser pulses is a rapidly developing area of research \cite{Kirilyuk2010}. Among the various mechanisms responsible for the excitation of such dynamics, the non-thermal ones are the most interesting \cite{Duong2004, Kimel2005, Hansteen2005, Hansteen2006, Perroni2006, Kalashnikova2007, Satoh2012, Au2013}. Using non-thermal excitation one is able to introduce changes in the magnetic system at very short time scales, which are defined by the spin-orbit coupling ($\sim$1-10~ps) and not by thermalization processes (10-1000~ps).
 
So far, two main types of non-thermal mechanisms were shown to exist. The first of them is characterized by an impulsive action, that only exists during the laser pulse. Inverse Faraday \cite{Kimel2005} and Cotton-Mouton \cite{Kalashnikova2007,Kalashnikova2008} effects (IFE \& ICME) are representative of this type. The second ones are displacive effects such as the photoinduced change of magnetic anisotropy (PIA) \cite{Stupakiewicz2001, Hansteen2006}, which persist in the sample for a time interval much longer than the length of the laser pulse. It has also been shown, that the combination of the two effects can in principle be used for ultrafast switching of the magnetization at the time scale of the laser pulse \cite{Hansteen2005}. Therefore, detailed understanding of the exact behavior of non-thermal excitation mechanisms is very important for further development of the ultrafast optical manipulation of magnetic moments.

In this paper, we carefully study the dependence of the induced magnetization dynamics in bismuth iron garnet on the polarization of the optical pump pulse as well as on the external applied magnetic field. Three different excitation mechanisms are distinguished. In addition to the impulsive IFE \cite{Kimel2005, Hansteen2005, Hansteen2006} and the displacive action of the PIA \cite{Stupakiewicz2001, Hansteen2005, Hansteen2006}, another impulsive photo-magnetic effect was discovered. This new photo-magnetic effect is linearly dependent on the light intensity but does not depend on polarization and acts during the presence of the light pulse only, or at least on a time scale much shorter than the precession period. This effect adds yet another possibility of all-optical control of magnetization.

The investigations have been performed using bismuth iron garnet (Bi$_3$Fe$_5$O$_{12}$, BIG). The interest in this material is caused by its largest known magneto-optical constants in the iron garnet family, with the Faraday rotation reaching 60~deg/$\mathrm{\mu m}$ in the visible light range ($\lambda=430$~nm). This property makes BIG a promising material for the fabrication, for example, of magneto-optical circulators \cite{Magdenko2010}. The synthesis of this material requires non-equilibrium growth techniques and, so far, the fabrication of a bulk crystal of BIG was not successful. However, since the '90s \cite{Satoh1990}, good quality thin films of BIG are grown on iso-structural substrates.

The studied sample is a 200~nm thick single crystalline and single phase BIG film grown epitaxially on a substituted $\mathrm{Gd}_{3}\mathrm{Ga}_{5}\mathrm{O}_{12}(001)$ substrate by pulsed laser deposition. The sample has uniaxial and cubic anisotropy which are of the order of 300~and 200~Oe respectively. The measured saturation magnetization is about 1500~Oe. For a detailed description of the growth conditions and the structural, magnetic and static magneto-optic properties of BIG films see Refs.\ \onlinecite{Popova2012, Deb2012, Popova2013, Deb2013}.

For the measurements an optical pump-probe setup in transmission geometry is used. The pump was the direct output of a Spectra Physics Spitfire amplified laser system giving 40~fs, 800~nm pulses at a repetition rate of 1~kHz. At this wavelength BIG is mostly transparent so heating effects are minimal. From the extinction coefficient of BIG in Ref.~\cite{Deb2012} and the heat capacity of yttrium iron garnet from Ref.~\cite{Hofmeister2006} (no data is available for BIG) the instantaneous temperature increase is calculated to be well below 1~K at the fluences used in this paper. For the probe, part of the laser output was directed through an optical parametric amplifier (OPA) to change the wavelength to 450~nm.

The pump pulse was aligned perpendicularly to the sample while the probe was at a small angle from the sample normal ($\sim$10~deg). The pump induced Faraday rotation of the probe was measured using a balanced detector scheme in combination with a lock-in amplifier and a chopper. An external in plane field was applied by an electromagnet. All measurements were performed at room temperature.

The probe polarization is in all cases linear while the polarization of the pump is varied between linear and circular. The spot size of the pump was, depending on the measurement, 130~or~365~$\mathrm{\mu m}$, while the probe spot was 26~$\mathrm{\mu m}$. The fluence of the pump was varied between 10 and 50~$\mathrm{mJ/cm^2}$. The pulse energy of the probe was at least $1000\times$ smaller than that of the pump.

All experimental data of the pump-induced temporal behavior of the Faraday rotation are fitted with an exponentially decaying sine function
\begin{equation}\label{EQ:FIT}
y=y_0+(B\mathrm{e}^{R_0t}) + A\mathrm{e}^{-t/\tau}\sin(2\pi ft-\phi).
\end{equation}
Here $A$ is the amplitude of the oscillations, $f$ is the frequency, $\tau$ is the oscillation lifetime, $\phi$ is the initial phase.

The origin of the offset in the data, as is described by the first two terms in Eq.~\ref{EQ:FIT}, we assign to a light induced change in the Faraday rotation of most likely electronic origin. Notice that the presence of such an offset is reported before \cite{Kimel2005, Kalashnikova2007} and as it is not expected to influence the conclusion of this paper it will not be further considered.

\begin{figure}
	\includegraphics{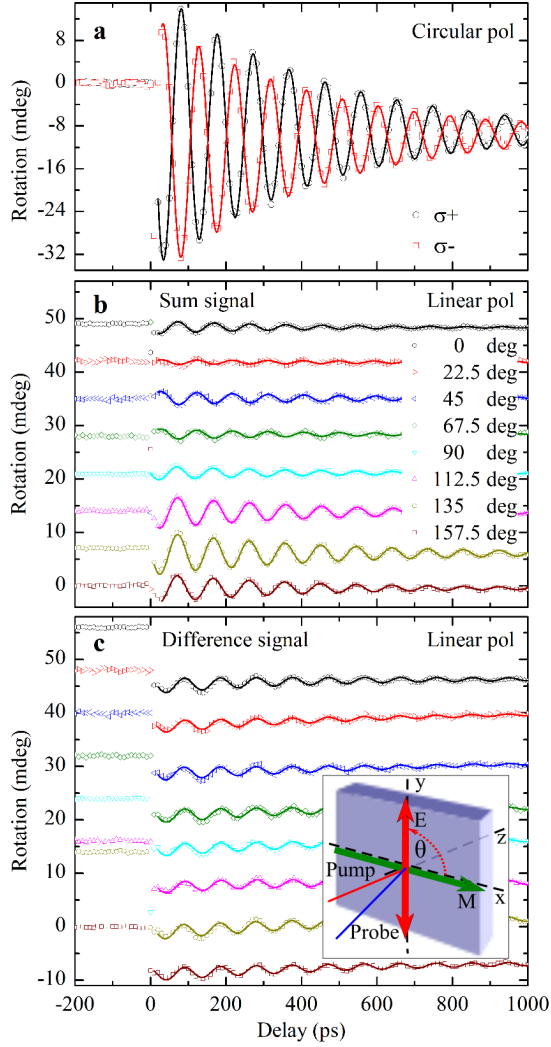}
 	\caption{\label{RawData}Observed magnetization dynamics for different pump polarizations. In (a) a 180~deg phase difference is visible between oscillations excited with right ($\sigma+$, black circles) and left ($\sigma-$, red squares) circular polarization. In (b) and (c) the data obtained for different linear pump polarization angles $\theta$ is shown. The sum(b)/difference(c) signal is obtained by taking the sum/difference of the positive and negative field data. The inset in (c) shows the definition of $\theta$. The solid lines are fits using Eq.~\eqref{EQ:FIT}. In all cases the external field is 3~kG and the pump fluence is 27~$\mathrm{mJ/cm^2}$.}
\end{figure}

The dynamics observed when we excite with circularly polarized pump pulses are shown in Fig.~\ref{RawData}(a). From this figure it is clear that when the helicity of the pump light is reversed from right ($\sigma+$) to left ($\sigma-$), the initial phase of the induced precession changes by 180~deg. This behaviour is similar to what is observed earlier \cite{Hansteen2005} and can be explained by the IFE. To be sure the observed oscillation are indeed due to the out of plane component we performed the measurements with different polarizations of the probe. No difference in dynamics was observed which confirms the oscillations are linked to the Faraday effect and not to the linear magnetic birefringence.

The much smaller oscillations obtained with linearly polarized pump light with different polarization angles $\theta$, are shown in Fig.~\ref{RawData}(b) and~(c). In those two figures we have plotted the sum and difference of the positive and negative field data. Representing the data in this manner shows that we can distinguish two differently behaving oscillations. First of all, the sum and difference signals represent oscillations that are independent, respectively dependent on the sign of the magnetic field.

\begin{figure}
	\includegraphics{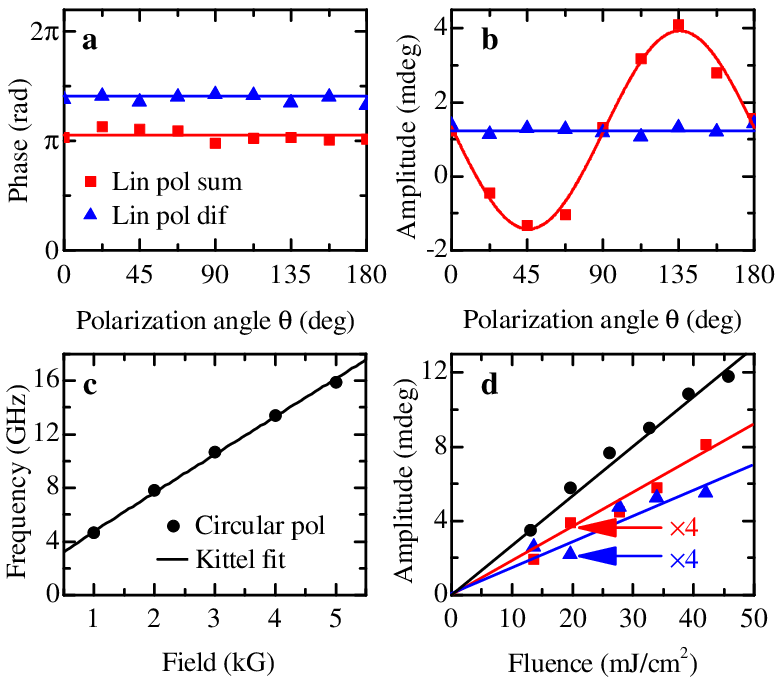}
 	\caption{\label{PolFreqFluence}Several characteristics of the observed magnetization dynamics. Initial phase (a) and oscillation amplitude (b) of the sum and difference data as a function of the linear polarization angle $\theta$. In (c) the precession frequency as function of the external field is shown for the circular polarization data. The frequency obtained from the sum and difference data are overlapping with the data points that are shown here. The dependence of the amplitude on the pump fluence is shown in (d) for circular (black) and linear (red for the sum and blue for the difference) polarized light. Except for the solid line in (c), which is a fit using the Kittel formula (Eq.~\eqref{EQ:KittelA}), the other solid lines are guides to the eye. In (a), (b) and (d) an external field of 3~kG was applied.}
\end{figure}

Furthermore, from Fig.~\ref{RawData}(b) and~(c) it is apparent that the sum and difference signal differ in their initial phase and in their amplitude dependence on $\theta$. In Fig.~\ref{PolFreqFluence}(a) the initial phase obtained from fits to the data shows a difference of almost $\pi/2$. While the sum signal is sine like, the difference signal is cosine like. If the sum signal is evaluated more carefully, a small change in phase might be present as a function of the polarization angle. However this change in phase can be neglected when compared to the difference between the sum and difference signal. Further, Fig.~\ref{PolFreqFluence}(b) shows that while the sum signal shows a periodic - $\sin 2 \theta$ - modulation of the amplitude, the amplitude of the difference signal does not change.

The behaviour of the sum signal is similar to what is observed in Ref.~\onlinecite{Hansteen2005} and can be ascribed to the PIA. The independence of the response on the field (magnetization) direction means that switching the direction of the magnetization leads to a reversal of the PIA contribution\cite{Hansteen2005}. In contrast, the difference signal in Fig.~\ref{RawData}(c), which shows polarization independent dynamics, is thus a totally different kind of excitation.

To better understand the differences and similarities between the three different types of excitations shown in Fig.~\ref{RawData} we measured their field and pump-fluence dependences. For the linear polarization the field dependence is measured at $\theta=135$~deg, while the fluence dependence is measured at $\theta=0$~deg. In Fig.~\ref{PolFreqFluence}(c) and (d) we show respectively the frequency versus field and the amplitude versus pump fluence. 

From the measured frequency dependence (Fig.~\ref{PolFreqFluence}(c)) on the external field we can conclude that for all three datasets the same ferromagnetic mode is excited. Furthermore in the experimental geometry that we use the Kittel formula can be written as \cite{Manuilov2010}
\begin{equation}\label{EQ:KittelA}
\omega=\gamma\sqrt{[H_{\mathrm{ext}}+(4\pi M_{\mathrm{s}}-H_{\mathrm{u}})+H_{\mathrm{c}}][H_{\mathrm{ext}}+H_{\mathrm{c}}]}.
\end{equation}
Here $\omega$ is the angular precession frequency, $\gamma$ the gyromagnetic ratio, $H_{\mathrm{ext}}$ the externally applied field and $H_{\mathrm{u}}$, $H_{\mathrm{c}}$ are the effective uniaxial and cubic anisotropy fields, respectively. Fitting Eq.~\eqref{EQ:KittelA} to the data in Fig.~\ref{PolFreqFluence}(c) gives us a value of 1192~Oe for $(4\pi M_{\mathrm{s}}-H_{\mathrm{u}})$ and a value of 200~Oe for $H_{\mathrm{c}}$. Unfortunately with Eq.~\eqref{EQ:KittelA} it is not possible to distinguish between $4\pi M_{\mathrm{s}}$ and $H_{\mathrm{u}}$. Using the value of 1500~G measured for the saturation magnetization, one can determine the uniaxial anisotropy field to be equal to $\sim$300~Oe. From Fig.~\ref{PolFreqFluence}(d) we see that the amplitude of all three datasets is approximately linear with the pump fluence.

The IFE and the PIA differ significantly by their respectively impulsive and displacive character, that can be illustrated by the field dependence of the precession amplitude. To explain this, using the Landau-Lifshitz equation,\cite{Landau1935}
\begin{equation}\label{EQ:LLG}
\frac{\mathrm{d}\vec{m}}{\mathrm{d}t}=\gamma(\vec{m}\times\vec{H}_{\mathrm{eff}}),
\end{equation}
we simulate the expected amplitude dependence on the external field for the observed oscillations.

The effective field, $H_{\mathrm{eff}}$ is given by
\begin{equation}\label{EQ:Heff}
\vec{H}_{\mathrm{eff}}=\vec{H}_{\mathrm{ext}}+\vec{H}_{\mathrm{ani}}+\vec{H}_{\mathrm{dem}}+H_{\mathrm{IFE}}.
\end{equation}
Here, $H_{\mathrm{ani}}$ is the anisotropy field, $H_{\mathrm{dem}}$ the demagnetizing field and $H_{\mathrm{IFE}}$ the effective field caused by the IFE. The latter is defined as
\begin{equation}\label{EQ:IFE}
\vec{H}_{\mathrm{IFE}}\propto\vec{E}\times\vec{E}^*,
\end{equation}
with $E$ the field amplitude of the light pulse. Eq. \eqref{EQ:IFE} implies that $\vec{H}_{\mathrm{IFE}}$ only exists during the presence of a light pulse.

From Eqs.~\eqref{EQ:LLG}~to~\eqref{EQ:IFE} it is clear that in the presence of a circularly polarized pump pulse the magnetization will start to precess in the sample plane and its final position is determined by the duration and intensity of the laser pulse. Because $\vec{H}_{\mathrm{IFE}} \perp \vec{m}$ the frequency of the initial precession is given by
\begin{equation}\label{EQ:Freq}
f=\frac{\gamma}{2\pi}|\vec{H}_{\mathrm{IFE}}|.
\end{equation}
Hence the final position is not influenced by the external field. After the laser pulse, a new precession will start around the $\vec{H}_{\mathrm{eff}}$ at that moment, thus with $H_{\mathrm{IFE}}=0$.

\begin{figure}
	\includegraphics{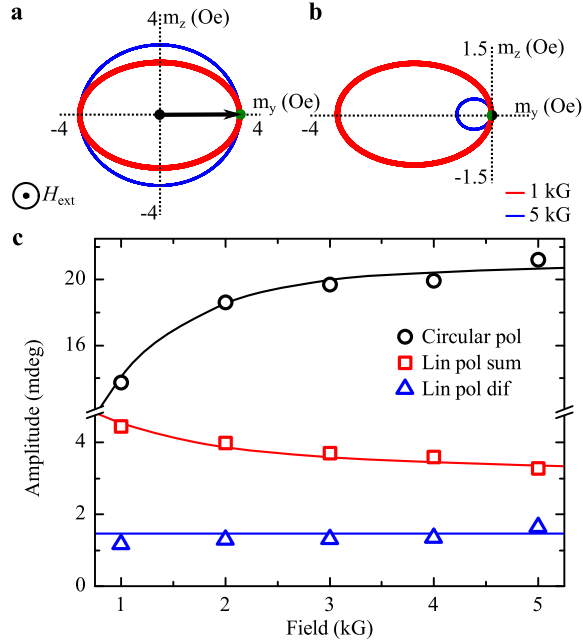}
 	\caption{\label{AmpVsField} The field dependent oscillation trajectories as obtained by Eq.~1 for oscillations initiated by the inverse Faraday effect(a) and by a photoinduced change in anisotropy(b). The black dot indicates the position of m just before arival of the pump pulse while the green dot indicates the position of m just after the pump pulse is gone. The experimental amplitude dependence on the external field is shown in (c) for a circular polarized pump pulse (black circles) and for the sum (red squares) and difference (blue triangles) signal of a linear polarized pump pulse. The solid lines are guides to the eye. The obtained error bars from the fitting procedure lie within the symbol size and are therefore not shown.}
\end{figure}

The experimental data in Fig.~\ref{RawData}(a) can be reproduced with Eq.~\eqref{EQ:LLG} by using for $H_{\mathrm{IFE}}$ a value of 3kOe. This value is of the same order of magnitude as found earlier \cite{Hansteen2006}. The trajectory described in this way is shown in Fig.~\ref{AmpVsField}(a). Due to the demagnetizing field it describes an elliptical, rather than a circular path. The long axis of the ellipse is oriented in plane. With increasing the external field the relative contribution of the demagnetizing field will decrease and thus the trajectory will more and more look like a circle, hence the absolute value of the out of plane component, which is measured in the experiment, will increase.

To reproduce the data in Fig.~1(b) with Eq.~\eqref{EQ:LLG} an in plane change in the anisotropy orthogonal to $\vec{m}$ of $\Delta H_{\mathrm{ani}}=1.3$~Oe is necessary. This value is also comparable to what is earlier reported for a PIA \cite{Hansteen2006}. Different from $H_{\mathrm{IFE}}$, $\Delta H_{\mathrm{ani}}$ does not only exist during the presence of the pump pulse but also after the light is gone \cite{Hansteen2005}. So in this case the magnetization starts to precess from its equilibrium position around a new effective field instead of a precession of the out of equilibrium magnetization around the unchanged effective field. 

Such a change in anisotropy will result in a decrease of the oscillation amplitude with increasing field \cite{Hansteen2006}. The contribution from the change in anisotropy to the effective field will become less relevant for stronger external fields. Contrary to excitation with the IFE the absolute opening angle of the precession cone decreases with increasing external field. The paths for two different external fields for this situation are shown in Fig.~\ref{AmpVsField}(b).

The experimentally observed oscillation amplitude versus external field is shown in Fig.~\ref{AmpVsField}(c). Remember that in the experiment we are mainly measuring the out of plane component of the magnetization. When excited with circular polarization the amplitude is increasing with field, as expected from the IFE. In contrast, the amplitude of the oscillations in the sum signal of the linearly polarized data is decreasing. As we assigned the origin of these oscillations to a PIA earlier, this is also what we expect. This field dependence excludes the possibility that the oscillations are initiated by the ICME \cite{Kalashnikova2007, Kalashnikova2008}.

Interestingly, the difference signal of the linear polarized data shows a constant or even slightly increasing amplitude with field. As a long living displacive effect will always result in a decrease of the absolute oscillation amplitude with increasing external field, the observed oscillations can only be excited with a mechanism that has an impulsive character. However, due to the absence of the polarization dependence, the oscillations cannot be ascribed to the known impulsive effects like the IFE and the ICME. So what can be the origin of these oscillations?

The initial phase and the impulsive charracter suggest that oscillations are induced by an opto-magnetic effect with an effective field in the sample plane, which direction is independend on the direction of $\vec{m}$. On a purely phenomenological basis, this effective field could be written as $H_{\mathrm{eff},i}=\chi_{ijk}E_{j}E_{k}^{*}$, where $\chi$ is a third rank axial c-tensor. Such field however should still change sign together with $\vec{m}$ via the time-reversal property of this tensor, and can thus be ruled out. 

The next best assumption will be either an out-of-plane PIA or an out-of-plane component of the ICME. Both of them could be schematically written as $H_{\mathrm{eff},i}=\chi_{ijkl}E_{j}E_{k}^{*}m_{l}$ \cite{Hansteen2005, Kalashnikova2007}, which however is not a strict definition for the PIA as discussed in \cite{Kirilyuk2010}. For both these effects the change in sign of $\vec{H}_{\mathrm{eff}}$ will lead to $\vec{m}$-dependent oscillations. The ICME results in a correct impulsive character, but predicts the initial phase to be different by about 75\degree. On the other hand, the PIA would result in an almost correct phase, but with a field dependence typical for a displacive effect. A compromise can be reached by a PIA with a life-time somewhat smaller than the precessional period: it will result in a semi-impulsive character of the amplitude, as is observed, but simultaneously will still posses a correct phase.

An effective field out of the sample plane induced by the in-plane components of the electric field suggests a rather low symmetry of the sample, where the properties are dominated by the out-of-plane direction \cite{Briss1964, Hansteen2005}. It has been shown by second harmonic generation (SHG) experiments \cite{Gridnev2001} that the epitaxial growth indeed leads to a symmetry breaking. Such breaking is expected to be much stronger in BIG that is not stable in the bulk phase. As a confirmation, we measured the non-linear optical response from our samples and found strong and isotropic SHG, indicating the dominating influence of the out-of-plane direction.

Although we are not able to reproduce the correct phase of the data in Fig.~\ref{RawData}(c) with Eq.~\eqref{EQ:LLG}, a correct amplitude is obtained by assuming a value of $H_{\mathrm{imp}}=1$~Oe for the polarization independent photo-magnetic effect when a lifetime of 10~ps is assumed. This value of $H_{\mathrm{imp}}$ is realistic when compared to the field for the displacive photo-magnetic effect $\Delta H_{\mathrm{ani}}$.

In conclusion, we have found a new, non-thermal and polarization independent impulsive or semi-impulsive photo-magnetic effect. This effect has been identified by a thorough analysis of the oscillation amplitude dependence on the magnetic field. For dynamics excited by a change in anisotropy, the amplitude is decreasing with increasing external field, while when excited by an impulsive action of the IFE the absolute amplitude will increase with increasing field or stay constant. The effective field connected to the new impulsive effect is found to be about 1~Oe in strength for a laser fluence of 27~$\mathrm{mJ/cm^2}$ and is directed out of the sample plane.

\begin{acknowledgments}
We would like to thank A. Toonen and A. van Roij for technical support.
This work was financially supported by de Nederlandse Organisatie voor Wetenschappelijk Onderzoek (NWO) and de Stichting voor Fundamenteel Onderzoek der Materie (FOM).
\end{acknowledgments}
%
\end{document}